\documentclass[manuscript]{acmart}

\AtBeginDocument{%
  }

\copyrightyear{2025}
\acmYear{2025}
\setcopyright{acmlicensed}\acmConference[ASSETS '25]{The 27th International ACM SIGACCESS Conference on Computers and Accessibility}{October 26--29, 2025}{Denver, CO, USA}
\acmBooktitle{The 27th International ACM SIGACCESS Conference on Computers and Accessibility (ASSETS '25), October 26--29, 2025, Denver, CO, USA}
\acmDOI{10.1145/3663547.3746384}
\acmISBN{979-8-4007-0676-9/2025/10}

\usepackage{graphicx,subcaption} 

\begin{document}

\title[Helping or Homogenizing?]{Helping or Homogenizing? GenAI as a Design Partner to Pre-Service SLPs for Just-in-Time Programming of AAC} 

\author{Cynthia Zastudil}
\email{cynthia.zastudil@temple.edu}
\affiliation{%
 \institution{Temple University}
 \streetaddress{1801 N. Broad St.}
 \city{Philadelphia}
 \state{Pennsylvania}
 \country{USA}
 \postcode{19122}
}

\author{Christine Holyfield}
\email{ceholfi@uark.edu}
\affiliation{%
\institution{University of Arkansas}
\streetaddress{1 University of Arkansas}
\city{Fayetteville}
\state{Arkansas}
\country{USA}
\postcode{72701}}

\author{Christine Kapp}
\email{christine.kapp@temple.edu}
\affiliation{%
 \institution{Temple University}
 \streetaddress{1801 N. Broad St.}
 \city{Philadelphia}
 \state{Pennsylvania}
 \country{USA}
 \postcode{19122}
}

\author{Kate Hamilton}
\email{kate.hamilton@temple.edu}
\affiliation{
 \institution{Temple University}
 \streetaddress{1801 N. Broad St.}
 \city{Philadelphia}
 \state{Pennsylvania}
 \country{USA}
 \postcode{19122}
}

\author{Kriti Baru}
\email{kriti.baru@temple.edu}
\affiliation{
 \institution{Temple University}
 \streetaddress{1801 N. Broad St.}
 \city{Philadelphia}
 \state{Pennsylvania}
 \country{USA}
 \postcode{19122}
}

\author{Liam Newsam}
\email{liamcnewsam@berkeley.edu}
\affiliation{
 \institution{UC Berkeley}
 \streetaddress{}
 \city{Berkeley}
 \state{California}
 \country{USA}
 \postcode{94720}
}

\author{June A. Smith}
\email{smithj7@berea.edu}
\affiliation{
 \institution{Berea College}
 \streetaddress{101 Chestnut St.}
 \city{Berea}
 \state{Kentucky}
 \country{USA}
 \postcode{40403}
}

\author{Stephen MacNeil}
\email{stephen.macneil@temple.edu}
\affiliation{%
 \institution{Temple University}
 \streetaddress{1801 N. Broad St.}
 \city{Philadelphia}
 \state{Pennsylvania}
 \country{USA}
 \postcode{19122}
}

\renewcommand{\shortauthors}{Zastudil et al.}

\begin{abstract}
    Augmentative and alternative communication (AAC) devices are used by many people around the world who experience difficulties in communicating verbally. One form of AAC device which is especially useful for minimally verbal autistic children in developing language and communication skills is the visual scene display (VSD). VSDs use images with interactive hotspots embedded in them to directly connect language to real-world contexts which are meaningful to the AAC user. While VSDs can effectively support emergent communicators (i.e., those who are beginning to learn how to use symbolic communication), their widespread adoption is impacted by how difficult these devices are to configure. We developed a prototype that uses generative AI to automatically suggest initial hotspots on an image to help non-experts efficiently create visual scene displays (VSDs). We conducted a within-subjects user study to understand how effective our prototype is in supporting non-expert users, specifically pre-service speech-language pathologists (SLPs) (N=16) who are not familiar with VSDs as an AAC intervention. Pre-service SLPs are actively studying to become clinically certified SLPs and have domain-specific knowledge about language and communication skill development. We evaluated the effectiveness of our prototype based on creation time, quality, and user confidence. We also analyzed the relevance and developmental appropriateness of the automatically generated hotspots and how often users interacted with (e.g., editing or deleting) the generated hotspots. Our results were mixed with SLPs becoming more efficient and confident. However, there were multiple negative impacts as well, including over-reliance and homogenization of communication options. 
    The implications of these findings reach beyond the domain of AAC, especially as generative AI becomes more prevalent across domains, including assistive technology. Future work is needed to further identify and address these risks associated with integrating generative AI into assistive technology.  
\end{abstract}



\received{16 April 2025}
\received[accepted]{18 June 2025}


\begin{teaserfigure}
  \includegraphics[width=\textwidth]{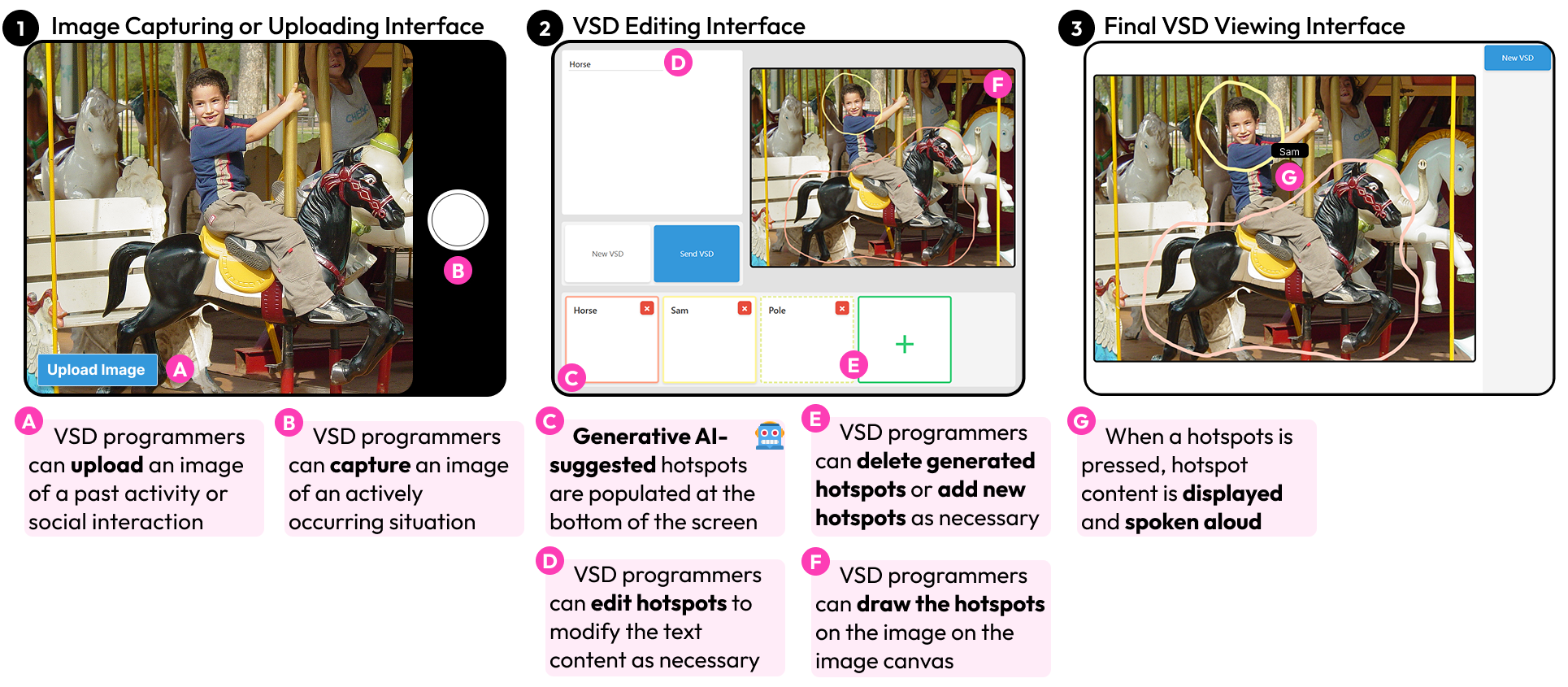}
  \caption{The three screens that comprise the user interface of our prototype. (1) VSD programmers choose to upload or capture an image to use in a VSD. (2) The application automatically generates a set of potential hotspots for use in the VSD and the programmer can choose to edit, use, or delete these hotspots. They can also manually add their own hotspots. Once all of the hotspots are created, they can use the canvas to draw the hotspots on the image. (3) Once they have finished configuring the VSD, users can see a preview of what the VSD looks like and interact with the created hotspots.}
  \Description{Displays the three interfaces of the prototype we developed for use on a tablet computer. The first interface is a image uploading or capturing screen. There is a button in the bottom left to upload an image from your device or a button to capture a photo using the device's camera on the middle-right of the screen. The second interface is the editing interface. Across the bottom of the screen, the automatic hotspot suggestions are displayed in a list of boxes. Each hotspot can be selected to either edit the text or draw the hotspot on the image. Each hotspot can be deleted using the 'X' icon in the top right of the hotspot box. At the end of the list of hotspots, users can press a button to manually create a new hotspot. In the top left corner of the screen, users can edit the text of hotspots they have selected. Below the editing box, there are two buttons. One to start over with a new VSD and one to preview the final VSD. In the top right corner, there is a canvas where users can draw hotspots on the image. When a hotspot has been drawn, the border of that hotspot box becomes solid, otherwise it is a dotted line. The final screen of the interface is the preview screen in which users can see the final VSD and interact with hotspots. When hotspots are pressed, a text overlay appears and the hotspot content is played aloud.}
  \label{fig:interfaces}
\end{teaserfigure}

\maketitle

\section{Introduction}

Many individuals with complex communication needs, such as those with autism, apraxia, or other speech and language disorders, rely on augmentative and alternative communication (AAC) to supplement or replace verbal communication in their daily lives. AAC often takes the form of technologies such as grid displays~\cite{light2019designing} or visual scene displays~\cite{blackstone2004visual}, which present communication options that are spoken when selected. Visual scene displays (VSDs) use images with embedded words or phrases in the form of selectable hotspots~\cite{blackstone2004visual}. VSDs are often used by emergent communicators (i.e., communicators who are working on learning and using symbolic language~\cite{dowden2004summary}) because they help to contextualize the communication options within a scene~\cite{light2012supporting,light2004performance}. This is especially helpful for young, minimally verbal autistic children~\footnote{In this paper, we primarily use identity first language (e.g., ``autistic children''), as recent research has found that many individuals prefer identity first language~\cite{sharif2022should}; however, this is not a universal preference and some people also prefer person-first language~\cite{taboas2023preferences} (e.g., ``children on the autism spectrum'').} who are still developing symbolic language skills and benefit from concrete, context-based representations of meaning~\cite{light2004performance,light2012supporting}. Similarly, VSDs group concepts together in an image taken from the real world, which preserves the relationships between people and objects~\cite{light2004performance}. As a result, VSDs can reduce the cognitive demands of using AAC by aligning with the natural visual processing of scenes~\cite{light2019designing}. These benefits distinguish VSDs from other AAC layouts including the more common grid displays that display symbols, which are often line drawings that do not as clearly reflect referents (i.e., the object, concept, or activity to which the symbol refers), displaying them in isolation without maintaining the relationship between words and the real world~\cite{mirenda1989comparison}.

Given how effective VSDs are with emergent communicators, they offer a promising solution for use outside of clinical contexts when SLPs are not present, such as in the home. Although they can be easier to use by communicators, VSDs are challenging to configure and often require experts to `program' them by choosing the hotspots for each image. There is not a universally agreed upon set of guidelines, so expertise is needed to effectively configure them. 
It can be incredibly time consuming to configure each image with relevant communication options, so VSDs are often configured ahead of time for a specific communication setting. This manual configuration requires frequent updates to remain relevant for their users' communication needs~\cite{drager2019aac}. Another way of configuring VSDs is just-in-time (JIT) programming~\cite{schlosser2016just} (i.e., capturing an image of an activity, social interaction, or object and configuring it in real-time for the VSD user). JIT programming supports communication partners in creating VSDs in the moment to take advantage of spontaneous opportunities for communication~\cite{schlosser2016just}. This approach still requires significant knowledge about how and when to program VSDs~\cite{holyfield2019programing}. Parents and other communication partners outside of clinical settings often lack this expertise, preventing widespread adoption of VSDs~\cite{hajjar2023visual, baxter2012barriers}.

Prior work has investigated the use of AI to generate high-quality, contextually relevant topic-specific communication boards~\cite{fontana2024co} and text-based communication suggestions~\cite{valencia2024compa}. We are interested in whether generative AI (genAI) can be used in AAC devices to reduce the effort required to program high-quality VSDs by scaffolding the VSD configuration process. While genAI may speed up the process, there are still some open questions. For example, it is vital that VSDs be programmed in a personalized way for each end user, taking into account their communication stage and personal interests and experiences~\cite{holyfield2019programing}. Prior work has shown that genAI does not do a good job at incorporating personalized information about end users without very detailed prompting~\cite{zastudil2024exploring}, which is not feasible for the just-in-time programming of VSDs. Rather than focusing on developing hyper-personalized user models which would be environmentally and financially expensive to build and would become outdated quickly, requiring frequent updating to stay up to date with users' communication needs, we are interested in ways to leverage communication partners' unique knowledge about VSD users to address the need for personalization.

In this paper, we investigate the impacts of integrating genAI into VSDs by designing and evaluating a prototype that uses genAI to scaffold the configuration of hotspots for VSDs. We evaluated this prototype with pre-service speech-language pathologists (SLPs) to understand the ways that it improves VSD configuration and also the potential negative consequences of integrating AI in this context. 
We conducted a within-subjects user study with our prototype and an existing VSD application as our control condition to investigate the impact our prototype has on VSD configuration done by pre-service SLPs. 



Given these methods and research goals, we investigate the following research questions:
\begin{itemize}
    \item[\textbf{RQ1:}] How effective, in terms of creation time, quality, and user confidence, is a genAI-assisted VSD application in supporting untrained communication partners?
    \item[\textbf{RQ2:}] Does a genAI-assisted VSD application provide relevant and developmentally appropriate hotspots?
    \item[\textbf{RQ3:}] What are the impacts of genAI suggestions on the content of VSD hotspots?
\end{itemize}

Based on an analysis of the configured VSDs and participants' confidence and perceptions of the automatically generated hotspots, our key findings are:
\begin{itemize}
    \item \textbf{VSDs were configured faster with increased confidence.} Participants were faster and more confident using our prototype with genAI suggestions.
    \item \textbf{Mixed results for VSD quality.} There were mixed results regarding the quality of the VSDs. While the VSDs configured with our prototype tended to use more developmentally appropriate hotspots, they also tended to contain too many hotspots, hotspots that were not directly related to the scene, and provided less hotspots for desired communication functions (e.g., socialization).
    \item \textbf{Increased homogeneity.} When using the VSD with generative AI suggestions, the resulting communication options were more semantically similar to the average VSD configured by participants.
    \item \textbf{Over-reliance on genAI suggestions.} Participants did not commonly delete or modify genAI suggestions, commonly accepting the majority of the suggestions provided to them with few edits or additions.
\end{itemize}

Our work makes the following contributions:
\begin{itemize}
    \item A prototype which leverages genAI to provide a starting point for VSD creation, which could be potentially useful in training settings for pre-service SLPs or other untrained communication partners.
    \item Empirical contributions measuring the efficiency, confidence, quality, and homogeneity of VSDs configured with genAI.
    \item Identification of implications for integrating genAI into AAC devices, including over-reliance on AI and increased homogenization of AAC displays.
\end{itemize}

Our prototype examines the integration of genAI into an existing state-of-the-art AAC device, to better understand what kinds of guardrails may be needed to address some of the negative impacts we identified of AI integration into AAC devices. In our discussion we talk about ways in which different HCI approaches may be able to address these issues. Across domains, researchers are advocating for designing tools with guardrails which incorporate best practices. For example, this is common in education~\cite{denny2024desirable, liffiton2024codehelp}, creativity research~\cite{shelby2024generative}, and data analysis research~\cite{ding2025diagram}.

\section{Related Work}

\subsection{Visual Scene Displays}
For people with complex communication needs, AAC intervention can aid in the development of language, literacy, communication, and cognitive skills~\cite{davidoff2017aac}. As an alternative to the traditional grid layout, visual scene displays (VSDs) are images of meaningful events or shared activities in which vocabulary is embedded within the image~\cite{blackstone2004visual}. For example, a photograph from a school dance could be used as a VSD with the faces of friends in the picture programmed to output their names or a picture of people dancing could be programmed to output ``dance''~\cite{holyfield2019effect}. 
This way, language is grounded within a familiar context for the AAC user due to its personally relevant meaning~\cite{holyfield2019effect}. VSDs are frequently used by emergent communicators (i.e., those who do not reliably communicate with symbolic language~\cite{dowden2004summary}), such as minimally verbal autistic children~\cite{gevarter2014comparing, ganz2015aac}. VSDs have also been found to be useful for people with aphasia~\cite{beukelman2015using, brock2017comparison}. 

VSDs can offset some of the cognitive load required to communicate messages using AAC devices for individuals with cognitive or linguistic challenges~\cite{light2019designing} because they maintain the relationship between people and objects as they experience them in real life~\cite{light2004performance}, represent social interactions through which language is learned~\cite{light2012supporting}, and reduce working memory demands because all people, objects, and activities are presented together~\cite{light2012supporting}. VSDs also have the potential to provide communication partners ways to get more involved in the communication process~\cite{blackstone2004visual} by working with AAC users to configure effective and engaging VSDs.

\subsection{Just-in-Time Programming}

To make AAC more contextually relevant in each moment, communication partners will often configure VSDs with language options right when they are needed. This technique is referred to as just-in-time (JIT) programming~\cite{schlosser2016just}. JIT programming allows communication partners to program new vocabulary in the moment based on events that unfold and interests demonstrated within them~\cite{holyfield2019programing}. 

JIT configuration allows clinicians to configure communication options for VSD users in real-time associated with an image or naturally occurring scene and it supports taking advantage of teachable moments as they occur~\cite{schlosser2016just}. However, the JIT programming approach raises questions regarding the risk of responding to false indicators of engagement or missing language learning opportunities~\cite{holyfield2019programing}. Similarly, despite the flexibility JIT configuration provides, there are still concerns as to whether clinicians can adjust accordingly when faced with environment changes or unforeseen situations~\cite{schlosser2016just}. It also requires clinicians to be present and continuously reprogramming the model to capture engaging scenarios~\cite{holyfield2019programing}. Automated JIT support may offer a simpler and more efficient approach to creating communication options that are relevant to an immediate context. 

Automated JIT scaffolding has been explored in other research domains including data visualization literacy~\cite{macneil2022expert}, explaining complex concepts in scientific research papers~\cite{head2021augmenting, august2023paper}, and in creativity research~\cite{macneil2021framing}. In an AAC context, prior work has investigated the use of JIT scaffolding through automatically generating topic-specific displays~\cite{fontana2024co}, providing text-based suggestions for starter phrases in conversations~\cite{valencia2024compa}, using communication partner speech as input to program a topic display JIT~\cite{holyfield2024leveraging}, and the impact of JIT programming on social participation of autistic children~\cite{holyfieldpreliminary}. There is also a significant amount of research in AAC dedicated to automation in AAC devices, including text prediction, abbreviation expansion, and keystroke automation~\cite{curtis2022state}.

\subsection{AI-Enabled AAC Devices}

Numerous studies have investigated whether and how AI can be integrated into AAC devices, including using computer vision algorithms to generate contextually relevant communication options based on a photograph~\cite{fontana2022aac, obiorah2021designing, mooney2018mobile}, gaze-based AAC~\cite{fiannaca2017aacrobat, zhang2017smartphone}, and automatic recognition of communication context (e.g., location or communication partner identity)~\cite{kane2012we}. Highly performant genAI provides new possibilities, and researchers have been exploring the ways in which it can be used to further enhance assistive technology. In the context of AAC devices, this includes automatically generating topic-specific displays~\cite{fontana2024co}, generating text to support conversations in text-based AAC devices~\cite{valencia2024compa, valencia2023less}, and tailoring AAC systems to reflect individual user narratives for more personalized and authentic communication in text-based AAC devices~\cite{pal2024empowering}.

Many researchers, ourselves included, as well as users of assistive technology have adopted a critical lens of using AI in assistive technology~\cite{mack2024they, alshaigy2024forgotten, giri2023exploring, glazko2023autoethnographic}. 
Many have argued that genAI promotes ableism and that outputs contain ableist biases and stereotypes~\cite{giri2023exploring, glazko2023autoethnographic, mack2024they, alshaigy2024forgotten}
and risks of misinformation and hallucinations in genAI responses~\cite{glazko2023autoethnographic}. GenAI should not be viewed as a panacea which can solve all issues with assistive technology. Rather, it is simply a tool which should be used with an abundance of caution. In our work, we do not intend for genAI to replace the role of speech-language pathologists or communication partners.



\section{A Prototype for AI-Assisted VSD Configuration}


The goal of this research is to understand how non-expert users interact with genAI-provided hotspot suggestions in the VSD configuration process and how their use of AI suggestions impacts their confidence and effectiveness of configuring VSDs as well as any potential quality impacts of AI suggestions.
To investigate this, we needed to design and develop a prototype. We developed a web application in ReactJS which enables users to upload an image, automatically generate AI suggested hotspots, and then review and modify them if needed (see Figure~\ref{fig:interfaces}). The prototype uses the computer vision capabilities of a multimodal large language model (LLM) to automatically generate hotspots which aid in scaffolding the VSD configuration process. The web application runs on a tablet and mimics an interface that is otherwise very similar to other VSD applications that VSD programmers and AAC users typically see. At this stage, our prototype supports the configuration of VSDs and the ability to preview the final VSD which an AAC user would use. This prototype is not meant for use with VSD end users.
In the following section, we outline the design goals which inspired the design of our prototype. Then, we describe the interfaces and key features of our prototype.

\subsection{Design Goals}

The design of our prototype was informed by best practices in VSD design. For example, VSDs should use hotspots which are engaging and personalized for the user~\cite{holyfield2019programing,light2019designing}, developmentally appropriate~\cite{holyfield2019programing}, and provide enough communication options to support language and communication skill development without overstimulating the end user~\cite{hajjar2023visual, thistle2021speech}. Our work is also informed by evidence that if AAC devices are easier to configure and use, they are less likely to be abandoned~\cite{mcnaughton2008child,baxter2012barriers}.
Our work is guided by these design goals: 

\textbf{Design Goal 1: Reduce the barriers to configuring VSDs.} VSDs are typically used by expert practitioners and researchers in a clinical context~\cite{thistle2021effectiveness}; 
to use VSDs outside of this relatively narrow context, scaffolding the process of mapping relevant and developmentally appropriate vocabulary~\cite{holyfield2019programing} needs to happen. Our goal, informed by prior research, is to include more communication partners who are programming VSDs and contexts of use~\cite{light2019new}. In order to support pre-service SLPs in VSD creation, we decided to use genAI in creating an initial set of hotspots to scaffold users' experiences when configuring VSDs.

\textbf{Design Goal 2: 
Make it easy to integrate VSD users’ linguistic abilities and interests.} 
While genAI has shown promise in AAC applications, there are drawbacks to incorporating genAI systems, such as the prevalence of harmful biases and stereotypes~\cite{mack2024they, glazko2023autoethnographic, weisz2024design, weidinger2021ethical, armstrong2024silicon} and hallucinations~\cite{glazko2023autoethnographic, weidinger2021ethical}. With these drawbacks in mind, communication partners should be included as moderators in the hotspot generation process to ensure that hotspots would not be harmful for VSD end users and that they would be developmentally appropriate and sufficiently personalized to be the most effective~\cite{zastudil2024exploring}. Prior research on VSD effectiveness shows that hotspots should be engaging for the end user~\cite{holyfield2019programing, light2019new, light2019designing}, and this is primarily achieved by including personally relevant hotspots about the people and activities being depicted in the image~\cite{holyfield2019programing, light2019new, light2019designing}.

\textbf{Design Goal 3: Minimize changes to the VSD users’ interface.} The VSDs used by emerging communicators should match the format of existing VSD applications as much as possible. This is important because learning to use AAC devices takes a lot of time and effort~\cite{mcnaughton2008child, rackensperger2005first, allyson2023experience}. Two popular VSD applications, Tobii Dynavox's Snap Scene\footnote{\href{https://us.tobiidynavox.com/products/snap-scene}{https://us.tobiidynavox.com/products/snap-scene}} and Attainment Company's GoVisual\footnote{\href{https://www.attainmentcompany.com/govisual}{https://www.attainmentcompany.com/govisual}}, both present VSDs with drawn hotspots encircling a person or object of interest, which when pressed, display a label with the hotspot name. We preserved this interaction pattern in our system.

\subsection{Features}

There are several key features of our prototype: automatic generation of hotspots, manually creating hotspots, modification of existing hotspots, and drawing hotspots on the selected image. 

\subsubsection{Create Hotspots Automatically}

In the past couple of years, there have been huge advancements in the capabilities of LLMs and within the last year multimodal LLMs which can accept text, audio, and image input have become available with impressive vision capabilities. 
We automatically generate hotspots using OpenAI's
\texttt{GPT-4o} model\footnote{\href{https://openai.com/index/gpt-4o-system-card/}{https://openai.com/index/gpt-4o-system-card/}}. We chose this model because at the time of our prototype's development, it was the most cost-effective and available pre-trained model. When an image is uploaded, our system sends the unannotated image to \texttt{GPT-4o} via its public API to automate the generation of hotspots from static visual media. The multimodal LLM processes the visual input and text-based prompt to return a ranked list of regions of ``hotspots''--areas of visual or contextual significance. We used the following prompt instructions for the language model:
\begin{quote}
    ``This photo was taken to be used in a visual scene display for a child. The child using the visual scene display uses mostly single words. Please provide the contextually relevant hotspots for the image if you are focused on building engagement in interactions and the emergence of words. Please focus primarily on the objects and activities being done in the scene. Include nouns, verbs, and descriptors when possible in words children would recognize.'' 
\end{quote}
Prior work by Zastudil et al. used a prompt which used information specific to the image provided to the language model and gave explicit instructions about the AAC user's goals when using the VSD~\cite{zastudil2024exploring}. Our prompt was informed by theirs, and we modified their prompt to improve generalizability so that any image could be used to configure a VSD. The modifications also better emphasize the importance of using appropriate vocabulary and producing contextually relevant hotspots which would be useful in a VSD. 

We chose to use a prompt without any personalized information (e.g., linguistic skills, names, special interests) about the VSD end users because it requires a lot of information about the user,  which has privacy considerations and is not easy to collect. Therefore, we leveraged current capabilities of the multimodal LLM to create contextualized suggestions about the people, objects, and activities being depicted in an image~\cite{zastudil2024exploring}. Personalization should come from the users configuring the VSDs. This reflects the unique knowledge SLPs have about their clients that they use when configuring AAC devices. Additionally, in the prompt, we did not give a limit to the number of hotspots, as our prototype is designed to scaffold VSD configuration for the users, providing a set of suggestions rather than a final set of hotspots.

\subsubsection{Create Hotspots Manually}


Users can \textbf{add new hotspots} which were not generated by our prototype. These hotspots can be whatever the user wants them to be. Supporting manual additions to the VSD provides users the flexibility to add hotspots beyond the automatically generated hotspots. The automatically generated hotspots may not include all of the hotspots related to the objects, activities, or people in the scene which the user may want to use. 

\subsubsection{Modify Existing Hotspots}

Once the hotspots are generated, users can \textbf{edit hotspots} directly in the web interface by clicking on the hotspot or \textbf{delete hotspots} which they deem irrelevant or simply do not want to use. We expected that users would edit hotspots to change the content to be more developmentally appropriate, personalized, or any other edits they would like to make.  

\subsubsection{Drawing Hotspots}
Once the user has decided which hotspots they want to use for the VSD, they are able to draw the hotspots on the selected image using their finger, a mouse, or a stylus.

\subsection{Interfaces}


There are three primary interfaces in our prototype. The image capturing and upload interface, VSD editing interface, and a VSD preview interface. Figure~\ref{fig:interfaces} shows what these interfaces look like. Our prototype was designed to be used on a tablet computer, such as an iPad or similar device.

\subsubsection{Image Capturing and Upload Interface}

When opening the application to configure a new VSD, the VSD programmer must first upload a new image. Users can upload an existing image or take a new image using the onboard camera. Currently, the uploaded image is not stored on the device or in the cloud.


\subsubsection{VSD Editing Interface}
Once the image is uploaded, the page for editing the VSD's hotspots is automatically opened. The editing interface consists of three windows: the list of generated hotspots, an editing window for hotspots, and the image for users to draw hotspots on. There are also options for starting over and creating a new VSD and viewing the finalized VSD an end user would see. Each generated hotspot can be deleted if the user does not wish to use it.
When a user selects a hotspot from the generated hotspots window, the editing interface is populated with the hotspot, where the user can then edit the text content of the hotspot. Additionally, users can choose to add hotspots by pressing the add hotspot button. 
Lastly, when the user selects a hotspot, they can draw the hotspot on the image populated in the image window.

\subsubsection{VSD Preview}
Once the user has decided to finalize the VSD, they are shown a preview of what an end user of the VSD would see. This includes the image with the hotspots they drew on the previous screen. When the hotspots are pressed, a label containing the hotspot text is temporarily overlaid over the image and the hotspot content is played aloud. In the current stage of this prototype, we do not send this VSD anywhere, it is solely for display purposes.

\section{User Study}


To evaluate our prototype and investigate our research questions, we conducted a within-subjects study consisting of two conditions for configuring VSDs. In the control condition, participants configured two VSDs using Tobii Dynavox's Snap Scene. In the experimental condition, participants used our prototype to complete the same task. The conditions were counterbalanced to address ordering effects, and four images were used to ensure reliability across stimuli.

Snap Scene was selected as a control condition because it is a widely available commercial VSD application. Snap Scene is available on more platforms than iOS, unlike similar VSD applications (e.g., GoVisual, Scene Speak\footnote{\href{https://www.goodkarmaapplications.com/scene-speak1.html}{https://www.goodkarmaapplications.com/scene-speak1.html}}),
and it only costs \$49.99. In addition to being widely used and available, it was also designed based on best practices and empirical research within the field which found it to be effective at facilitating social interactions, sharing information and expressing needs, integrating new words and concepts, and combining words for more complex ideas~\cite{holyfield2019effect}.
 

\subsection{Participants}

VSDs have a lot of potential benefits for AAC users, especially for beginning communicators~\cite{light2004performance, light2012supporting}. However, the lack of training and knowledge of effective configuration of VSDs hinders their more widespread adoption~\cite{kovacs2021survey}. While experts in VSDs have the knowledge for configuring them, untrained communication partners may need additional scaffolding in configuring effective VSDs. We recruited seventeen pre-service speech-language pathologists (SLPs) from a large R1 university in the United States.
We recruited participants using convenience sampling via an email sent to students studying communication sciences and disorders. Our inclusion criteria is that they must be studying to become a clinically certified SLP (CCC-SLP) and that they have some clinical experience. Based on responses to demographic survey questions, only two participants reported frequently using or observing VSDs being used in a clinical setting. Therefore, participants could be considered mostly unfamiliar with VSDs. Additionally, none of our participants had used the control software before our study. See Table~\ref{tab:demographics} for full participant information.


\begin{table}
\centering
\begin{tabular}{@{}llllll@{}}
\toprule
& \begin{tabular}[c]{@{}l@{}}\textbf{Program}\\ \textbf{and Year}\end{tabular} & \begin{tabular}[c]{@{}l@{}}\textbf{Clinical}\\ \textbf{Experience}\end{tabular} & \textbf{Client Ages} & \begin{tabular}[c]{@{}l@{}}\textbf{Emerging}\\ \textbf{Communicators}\end{tabular} & \begin{tabular}[c]{@{}l@{}}\textbf{VSD}\\ \textbf{Experience}\end{tabular} \\ \midrule
\textbf{P1} & 2nd, GR & School & \begin{tabular}[t]{@{}l@{}}Young children, \\ school-age children,\\ transition-age youth, adults\end{tabular} & Occasionally & Never \\
\textbf{P2} & 2nd, GR & \begin{tabular}[t]{@{}l@{}}School, \\ private therapy\end{tabular} & \begin{tabular}[t]{@{}l@{}}Young children, \\ school-age children,\\ transition-age youth\end{tabular} & Frequently & Occasionally \\
\textbf{P3} & 2nd, GR & Hospital & Adults & Very rarely & Never \\
\textbf{P4} & 2nd, GR & \begin{tabular}[t]{@{}l@{}}Hospital,\\ private therapy\end{tabular} & \begin{tabular}[t]{@{}l@{}}School-age children, \\ transition-age youth,\\ adults\end{tabular} & Very rarely & Rarely \\
\textbf{P5} & 2nd, GR & \begin{tabular}[t]{@{}l@{}}School,\\ private therapy\end{tabular} & \begin{tabular}[t]{@{}l@{}}Young children, \\ school-age children\end{tabular} & Occasionally & Occasionally \\
\textbf{P6} & 2nd, GR & \begin{tabular}[t]{@{}l@{}}School,\\ private therapy\end{tabular} & \begin{tabular}[t]{@{}l@{}}Young children, \\ school-age children\end{tabular} & Very frequently & Frequently \\
\textbf{P7} & 2nd, GR & \begin{tabular}[t]{@{}l@{}}School,\\ private therapy\end{tabular} & \begin{tabular}[t]{@{}l@{}}Young children, \\ school-age children,\\ transition-age youth\end{tabular} & Frequently & Occasionally \\
\textbf{P8} & 4th, UG & \begin{tabular}[t]{@{}l@{}}School,\\ private therapy\end{tabular} & \begin{tabular}[t]{@{}l@{}}Young children, \\ school-age children\end{tabular} & Occasionally & Occasionally \\
\textbf{P9} & 3rd, UG & School & Young children & Never & Very rarely \\
\textbf{P10} & 3rd, UG & School & Young children & Occasionally & Never \\
\textbf{P11} & 4th, UG & \begin{tabular}[t]{@{}l@{}}School,\\ private therapy\end{tabular} & \begin{tabular}[t]{@{}l@{}}Young children, \\ school-age children,\\ transition-age youth, adults\end{tabular} & Frequently & Frequently \\
\textbf{P12} & 3rd, UG & Summer camp & Young children & Occasionally & Never \\
\textbf{P13} & 1st, GR & Private therapy & \begin{tabular}[t]{@{}l@{}}Young children, \\ school-age children\end{tabular} & Occasionally & Occasionally \\
\textbf{P14} & 1st, GR & Hospital & Adults & Very rarely & Never \\
\textbf{P15} & 1st, GR & Hospital & Adults & Never & Never \\
\textbf{P16} & 2nd, GR & School & School-age children & Frequently & Occasionally \\
\textbf{P17} & 2nd, GR & \begin{tabular}[t]{@{}l@{}}School,\\ private therapy\end{tabular} & \begin{tabular}[t]{@{}l@{}}Young children, \\ school-age children,\\ transition-age youth, adults\end{tabular} & Frequently & Rarely \\ \bottomrule
\end{tabular}
\caption{An overview of the demographic information for the 17 pre-service SLPs that participated in our study. We collected data about the number of years they had been in their program, the settings of their clinical experience, the age ranges of their clients, whether they work with emerging communicators, and whether they have either used or observed the use of a VSD. For the program year, we report which year they are in the program and whether they are undergraduate students (UG) or graduate students (GR).}
\label{tab:demographics}
\end{table}

\subsection{Procedure}
The second author of this work, a clinically certified SLP and expert on VSDs, ran each user study in-person during September 2024. It consisted of three phases (1) information about VSDs, (2) VSD configuration with Snap Scene and our prototype, and (3) an online post-questionnaire. We received ethics approval to run this study by our institutional review board (IRB) before running the study. Participants signed a consent form before participating in our study. Participation was voluntary. After completion of the study, participants received information related to programming AAC for emerging symbolic communicators that could be useful to them in their future careers.



\subsubsection{Information about VSDs}
Before beginning the configuration portion of the user study, since our participants were largely unfamiliar with VSDs, we provided participants with a broad overview of VSDs, including how they are typically used, how images are selected for VSDs, and what functions hotspots perform. This included the second author providing verbal information along with some accompanying images of three example VSDs to demonstrate how they are used. After we provided them with this information, they were given the opportunity to ask any questions they had about VSDs to the researcher running the user study.

\subsubsection{VSD Configuration}
Once the participants had information about VSDs, we had participants configure VSDs with the control application (Snap Scene) and our prototype. The order of which application was used first and which images were used was counterbalanced across the entire experiment to mitigate any order effect. For each application, we showed participants a brief video tutorial to illustrate how each application worked. Participants configured VSDs for two contexts: playing and retelling a past activity. These contexts were selected because they are common use cases for VSDs~\cite{laubscher2019effect, chapin2022effects}. For each application, we provided the participant with two images (one for each context), each with a case study to inform their VSD creation. Case studies are commonly used in AAC research to describe AAC users’ language capabilities, needs, and goals~\cite{holyfield2019effect}. In order to ensure that VSD configuration was as naturalistic as possible to reflect how SLPs currently use VSDs in educational or therapeutic contexts, and how that behavior might change if they are assisted by AI, we did not inform participants about any assessment criteria (see Section~\ref{sec:data-analysis}) we used in our analysis to avoid biasing their use of either system. An example image and case study provided to participants in shown in Figure~\ref{fig:sample-prompt}. All images and case studies we used are provided in Appendix~\ref{sec:all-prompts}. 

\begin{figure*}
    \centering
    \includegraphics[width=0.9\linewidth]{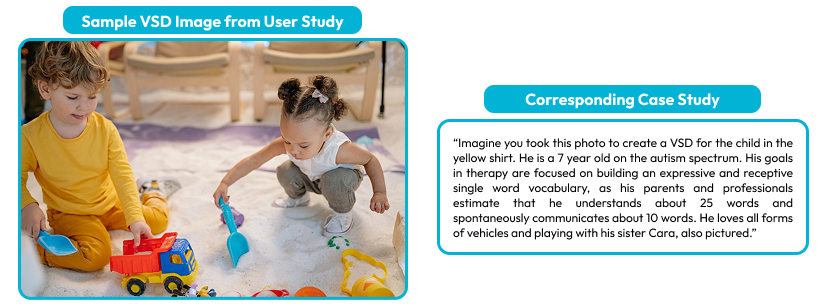}
    \Description[Sample VSD image and case study for participants in user study.]{On the left, a sample VSD image we had participants make VSDs for is provided. It depicts two young children playing in a sandbox with a variety of sand toys, including a dump truck and shovels. On the right, the case study which accompanied the image is provided. It reads, "Imagine you took this photo to create a VSD for the child in the yellow shirt. He is a 7 year old on the autism spectrum. His goals in therapy are focused on building on expressive and receptive single word vocabulary, as his parents and professionals estimate that he understands about 25 words and spontaneously communicates about 10 words. He loves all forms of vehicles and playing with his sister Cara, also pictured."}
    \caption{An example VSD image and case study prompt participants used during VSD creation. We provided information related to the hypothetical VSD user's linguistic abilities, communication goals, and personal details (e.g., special interests).}
    \label{fig:sample-prompt}
\end{figure*}

\subsubsection{Post-Questionnaire}
Once participants finished configuring their VSDs, we had them complete a post-questionnaire. Participants were asked in the questionnaire to rate their confidence in the VSDs they made with both Snap Scene and our prototype using a Likert-scale from ``Not at all Confident'' to ``Extremely Confident''. We then asked participants questions about the relevance and appropriateness of the generated vocabulary based off of questionnaires used by de Vargas et al.~\cite{fontana2024co, fontana2022aac}. The full post-questionnaire is provided in Appendix~\ref{sec:post-questionnaire}.

\subsection{Data Collection}
We collected screen recordings of the VSD configuration process. We used these recordings to obtain information about how long VSD creation took and which hotspots were created. Two members of the research team reviewed each of the videos carefully capturing timestamps for the starting and ending times for each VSD and what hotspots were created. For the VSDs configured during the prototype condition, we also recorded which generated hotspots were used, edited, or deleted. Each video was reviewed by two members of the research team to ensure there were no mistakes in data collection. Additionally, we used the responses to the post-questionnaire to gain insights on participants' confidence when making their VSDs and their perceptions on the quality of the generated hotspots. 

\subsection{Data Analysis}
\label{sec:data-analysis}
To analyze the results of our user study we conducted multiple levels of analysis, both at the hotspot and VSDlevel. Our hotspot-level analysis consisted of analyzing the content of the hotspots using part-of-speech analysis~\cite{zastudil2024exploring} and deductive coding using Light's functions of communication framework~\cite{light1988interaction,zastudil2024exploring}. We also analyzed what automatically generated hotspots participants used, modified, deleted, and added when using our prototype. At the VSD-level we analyzed the quality of the configured VSDs based off best practices for VSD creation informed by prior research. We also conducted a homogenization analysis of the configured VSDs.

\subsubsection{Hotspot Analysis}
We analyzed the content of the hotspots across both conditions to understand how AI suggestions might have impacted the VSDs participants configured. This included part-of-speech analysis and deductive coding to determine to primary functions of the hotspots. The part-of-speech analysis and deductive coding allowed us to compare the hotspots to understand if and how hotspot content differed by condition. Additionally, we analyzed the interaction logs for the prototype to understand which hotspots were added, edited, and deleted from the VSDs. We did not conduct analysis of the generated hotspots as this has been done in prior work~\cite{zastudil2024exploring}. Our analysis was focused on understanding how participants used the generated suggestions in combination with the knowledge they had about the hypothetical VSD end users they were configuring based on the case studies provided. 

For the part-of-speech analysis we followed the procedure outlined by Zastudil et al.~\cite{zastudil2024exploring}. Each hotspot was split into single words to ensure that every part-of-speech was accounted for. The first author then classified each word by its part-of-speech. We conducted this analysis to determine how the content and structure of the hotspots aligned across the Snap Scene and prototype VSDs.

We conducted the same deductive coding process outlined by Zastudil et al.~\cite{zastudil2024exploring} for the hotspots made with Snap Scene and with our prototype. These deductive codes are from Light's Functions of Communication framework~\cite{light1988interaction}:
\begin{enumerate}
    \item \textbf{Expressing Wants or Needs} - communication intended to make requests
    \item \textbf{Information Transfer} - communication meant to share information with others
    \item \textbf{Social Closeness} - communication meant to develop or maintain relationships
   \item \textbf{Social Etiquette} - communication meant to convey polite terms (e.g., ``thank you'', ``please'', ``hello'')
\end{enumerate}
We also included an ``Other'' category to handle communication options which did not clearly align with these four functions. Two researchers performed the coding and inter-rater reliability was computed in accordance with best practices for qualitative research~\cite{mcdonald2019reliability}. We computed the inter-rater reliability using Cohen’s Kappa since we had two raters and categorical codes~\cite{cohen1960coefficient}. The inter-rater reliability score was 0.66 indicating substantial agreement~\cite{landis1977application}.


\subsubsection{VSD Analysis}
\label{sec:quality}
Evaluating the quality of hotspots within a VSD is challenging because it is a highly subjective task. 
There is not an agreed upon set of guidelines for creating VSDs; however, there are recommendations for VSD configuration informed by prior research. We analyzed the quality of the VSDs configured by participants according to the guidelines described below:
\begin{enumerate}
    \item Use 2--4 hotspots to ensure that there isn't too much visual stimulation or overlapping hotspots which may overwhelm users~\cite{hajjar2023visual, thistle2021speech} or result in stimulus over-selectivity~\cite{dube2014potential, wilkinson2022judicious} (i.e., hyper-attentiveness to some stimuli and limited to no attention paid to other relevant stimuli~\cite{bradshaw2013stimulus}).
    \item Focus hotspots on the people, activities, and social interactions (when applicable) in the scene~\cite{wilkinson2012considerations, holyfield2019programing}.
    \item Use hotspots which align with the users' communication stage and learning goals~\cite{holyfield2019programing, light2019new}. This includes using single words in hotspots over phrases which can be combined together to create more complex phrases~\cite{holyfield2019programing, light2019new}. 
\end{enumerate}

Additionally, we analyzed the VSDs configured to see if the genAI-assistance resulted in more homogeneous VSDs by using the homogenization analysis procedure developed by Anderson et al.~\cite{anderson2024homogenization}. 
We computed the sentence embeddings~\cite{reimers2019sentence} for all of the sets of hotspots made across conditions using the Python SentenceTransformers library~\cite{reimers2019sentence} with the model \texttt{all-MiniLM-L6-v2}. We then calculated the average embedding for all hotspots across conditions. Then, we compared each set of hotspots to the average embedding using cosine similarity to understand how distinct each set of hotspots was from the group. 

\section{Results}

We collected a total of 64 VSDs (32 from Snap Scene and 32 from our prototype) with 247 hotspots (94 from Snap Scene and 153 from our prototype). A total of 178 hotspots were generated by our prototype, and 118 of those were used either unmodified or they were edited. One participant's data (P5) was removed from the study due to a application error with the control software. All of the VSDs participants configured for both conditions are provided in our supplemental material.

\subsection{RQ1: Effectiveness of Generative AI-Assisted VSD Creation}
We found that participants configured VSDs more efficiently in terms of time and that they felt more confident about the VSDs they configured with our prototype than the VSDs they configured with Snap Scene. Overall, we saw that participants used more hotspots when creating VSDs with our prototype than with Snap Scene. 
The results of our quality analysis were mixed, with the VSDs configured with Snap Scene adhering to two out of three best practices, whereas our the VSDs configured with our prototype only adhered to one out of three.

\subsubsection{Time-to-Create VSDs Across Conditions}
\begin{figure}
    \centering
    \includegraphics[width=0.5\linewidth]{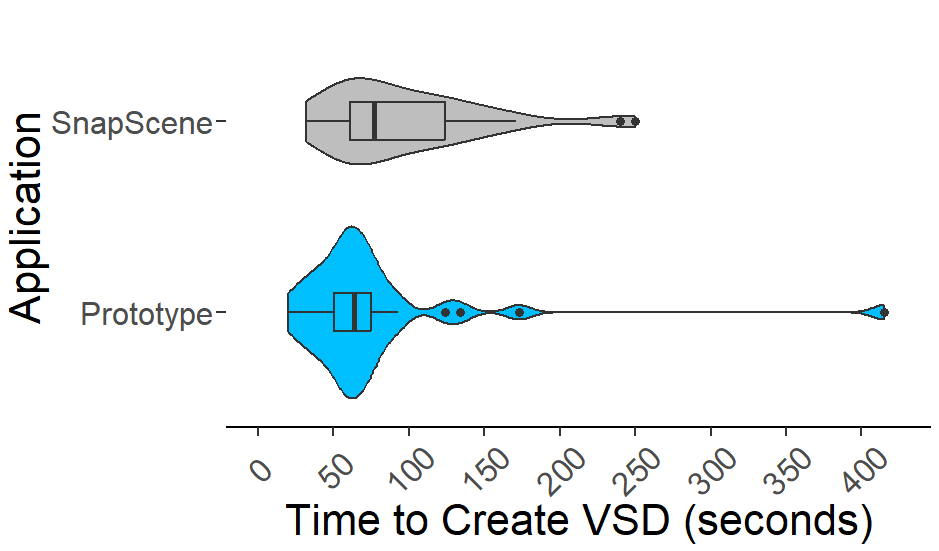}
    \Description[Stacked violin chart visualization describing time for creation across conditions.]{A stacked violin chart visualization with overlaid boxplots describing the distribution of the time it took for participants to configure VSDs using the control software (Snap Scene), shown on top in gray, and our prototype, shown on the bottom in blue. The median time for VSD creation using Snap Scene was 77 seconds and the median for our prototype was 63.5 seconds. The minimum times for creation using Snap Scene and our prototype were 32 seconds and 20 seconds, respectively. The maximum times for creation using Snap Scene and our prototype were 171 seconds and 93 seconds, respectively. There were two outliers for the Snap Scene condition 240 and 250 seconds. There were four outliers for the prototype condition, 124, 134, 173, and 415 seconds.}
    \caption{On average, participants configured VSDs faster when using our prototype compared to when they used Snap Scene.}
    \label{fig:ttc}
\end{figure}

Participants' median time to configure a VSD (i.e., time-to-create) VSDs when using Snap Scene and our prototype was 77 seconds and 63.5 seconds respectively. A Wilcoxon Signed-rank test shows that there is a statistically significant effect on the prototype on participants' time-to-create when creating VSDs ($W = 438.5, Z = 3.2635, p < 0.05$) with a medium effect size ($r = 0.41$). See Figure~\ref{fig:ttc} for a visual comparison of participants' time-to-create across conditions.

\subsubsection{Confidence in Configured VSDs}

\begin{figure}
    \centering
    \includegraphics[width=0.5\linewidth]{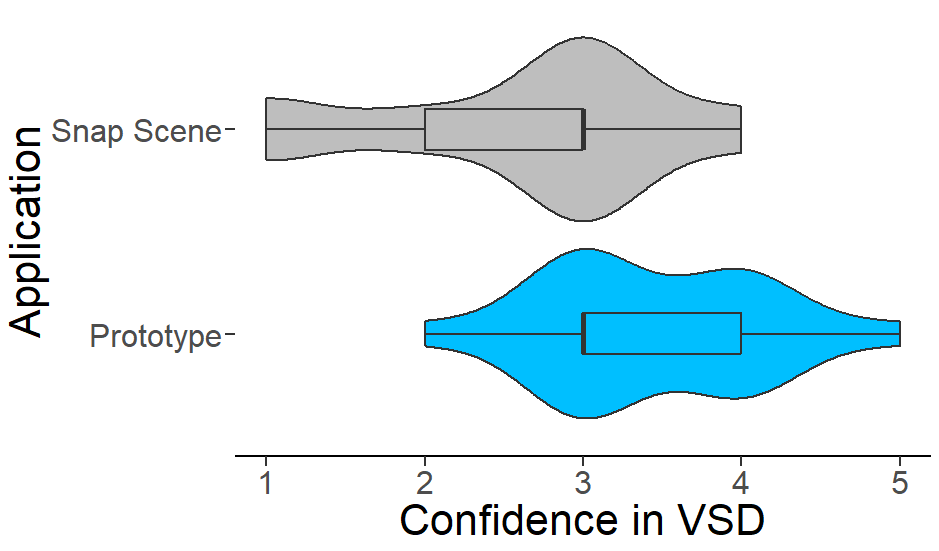}
    \Description[Stacked violin chart visualization describing participants' self-reported confidence in the VSDs they made across condition.]{A stacked violin chart visualization with overlaid boxplots describing participants' self-reported confidence in the VSDs they made across condition. Participants responded to two Likert-scale questions from "Not at all confident" to "Extremely confident" (from 1 to 5). The Snap Scene condition is displayed on the top in gray. The prototype condition is displayed on the bottom in blue. The median confidence for both conditions was 3. The maximum confidence for Snap Scene and the prototype were 4 and 5, respectively. The minimum confidence for Snap Scene and prototype were 1 and 2, respectively. There were no outliers.}
    \caption{Participants' reported their confidence levels (on a 5-point Likert scale). We found that their confidence levels were higher for the VSDs they configured using our prototype compared to their reported confidence when configuring VSDs with Snap Scene.}
    \label{fig:confidence}
\end{figure}

Participants ranked their confidence in the VSDs they configured along a 5-point Likert scale. A Wilcoxon Signed-rank test shows that there is a significant effect of the prototype on users' confidence when creating visual scene displays ($W = 5, Z = -2.8043, p < 0.05$) with a medium effect size ($r = 0.496)$. Figure~\ref{fig:confidence} provides a visual comparison of participants' confidence across conditions.


\subsubsection{Number of Hotspots Made Across Conditions}

\begin{figure}
    \centering
    \includegraphics[width=0.5\linewidth]{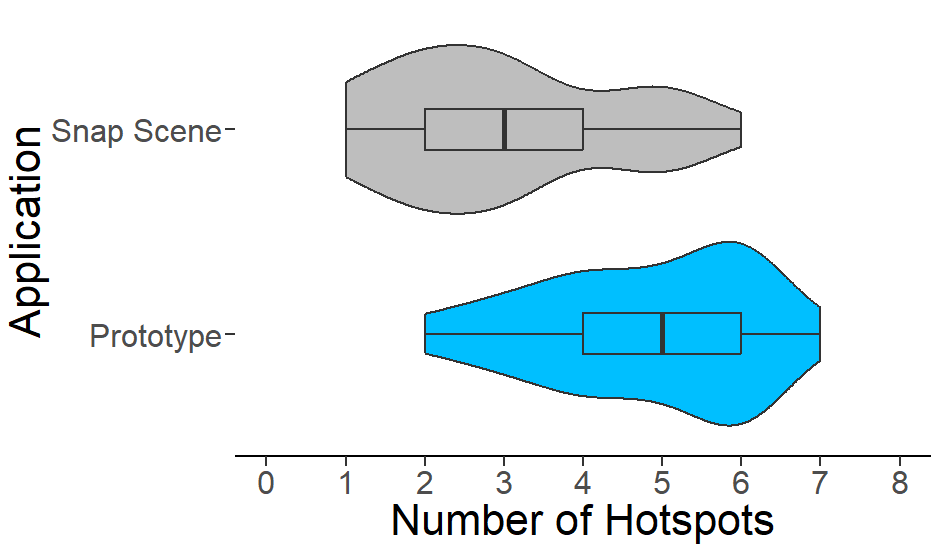}
    \Description[Stacked violin chart visualization describing the number of hotspots participants used per VSD across conditions.]{A stacked violin chart with overlaid boxplot visualization describing the number of hotspots participants used per VSD across conditions. The median number of hotspots per VSD for Snap Scene was 3. The median number of hotspots per VSD for the prototype was 5. The minimum number of hotspots used for VSDs in the Snap Scene and prototype conditions were 1 and , respectively. The maximum number of hotspots used for VSDs in the Snap Scene and prototype conditions were 6 and 7, respectively. There were no outliers.}
    \caption{Participants tended to use more hotspots when programming VSDs with our prototype than when using Snap Scene.}
    \label{fig:hotspots-used}
\end{figure}

Participants used, on average, 2 more hotspots per VSD when using our prototype (n=153) versus using Snap Scene (n=94). The median  number of hotspots when using Snap Scene and our prototype were 3 and 5, respectively. A Wilcoxon Signed-rank test shows that there is a significant effect of the prototype on users' number of hotspots used when creating visual scene displays ($W = 5.5, Z = -4.5567, p < 0.05$) with a large effect size ($r = 0.57$). See Figure~\ref{fig:hotspots-used} for a visual comparison of participants' number of hotspots used per VSD.

\subsubsection{Quality of VSDs Configured}

Overall, we see that findings for quality were mixed. Our quality measures are outlined in Section~\ref{sec:quality}. For quality measures 1 and 2, Snap Scene outperformed our prototype, but for quality measure 3 the prototype outperformed Snap Scene. 
%
Quality measure 1 is focused on the number of hotspots used in VSDs. In the Snap Scene condition, 60.50\% of VSDs used between 2 and 4 hotspots. In the prototype condition, we saw that 40.63\% of VSDs used between 2 and 4 hotspots. Quality measure 2 is focused on the relevance of the hotspots used to the scene. In the Snap Scene condition, we saw that 95.74\% of hotspots were focused on the people, activities, or social interactions, whereas in the prototype condition we saw that 84.97\% of hotspots were focused on the people, activities, or social interactions. Quality measure 3 is focused on the developmental appropriateness of the hotspots used. In the Snap Scene condition participants used single-word hotspots 78.70\% of the time, whereas in the prototype condition, participants used single-word hotspots 91.50\% of the time.

\subsection{RQ2: Relevance and Appropriateness of Generated Hotspots}
Participants felt that the generated hotspots were relevant and appropriate for the targeted goals outlined for their VSD creation in the case studies we provided them. But there were trends in the modifications made, including removing irrelevant hotspots and editing hotspots to be more personalized or developmentally appropriate.

\subsubsection{Participants' Perceptions of Relevance and Appropriateness}
\begin{figure*}
    \centering
    \includegraphics[width=0.9\linewidth]{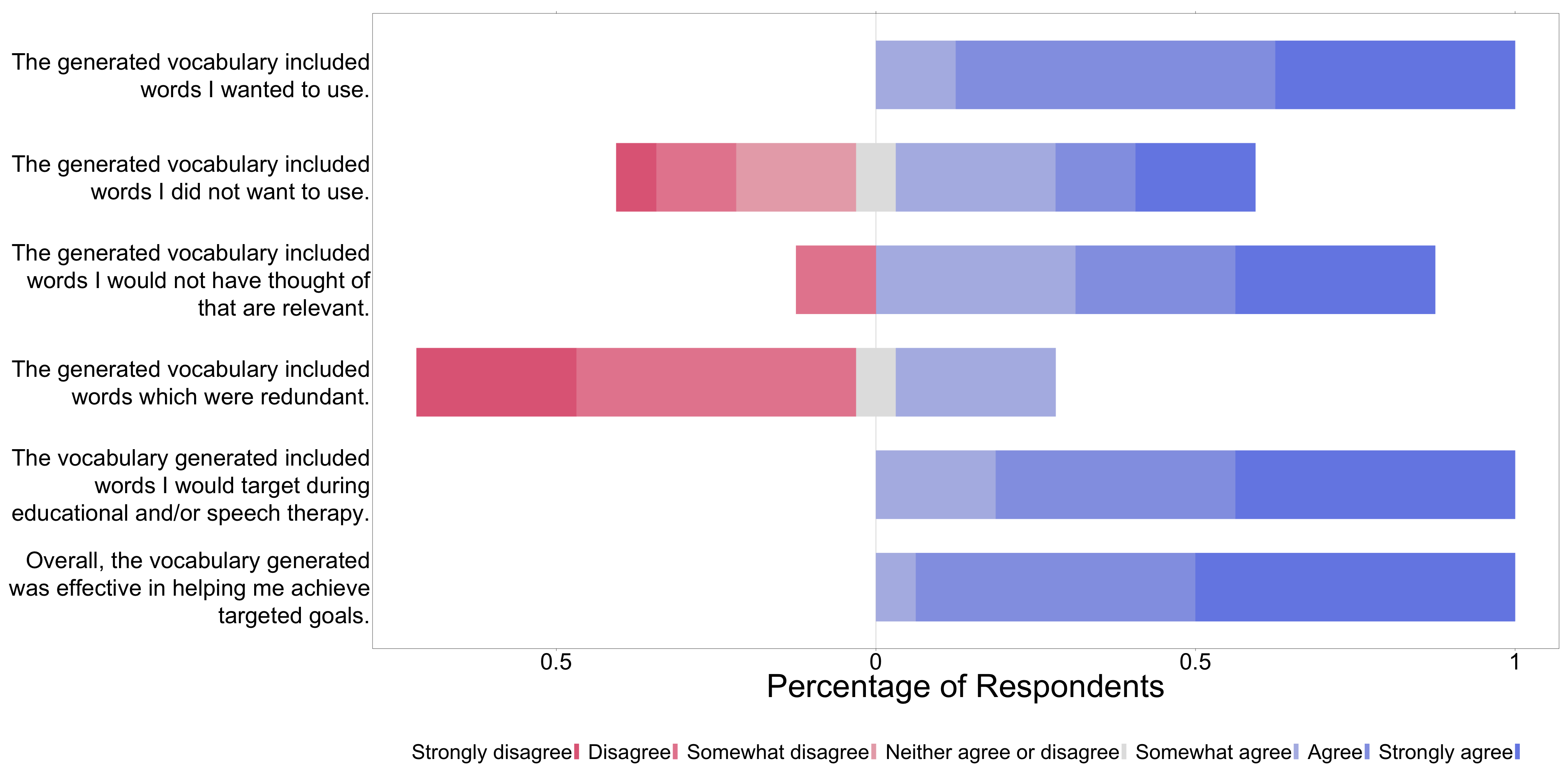}
    \Description[Horizontally stacked bar chart describing participants' responses to a Likert-scale questionnaire.]{A horizontally stacked bar chart describing participants' responses to our Likert-scale post-questionnaire. All questions were asked on a 7-point scale from "Strongly disagree" to "Strongly agree". For each question the percentage of respondents who provided each answer is provided. All of the participants responded "Somewhat agree" to "Strongly agree" for the question, "The generated vocabulary included words I wanted to use." More than half of participants responded "Somewhat agree" to "Strongly agree" and approximately 40\% responded "Strongly disagree" to "Somewhat disagree" for the question, "The generated vocabulary included words I did not want to use". The majority of participants responded "Somewhat agree" to "Strongly agree" to the question "The generated vocabulary included words I wuld not have thought of that are relevant. The majority of participants responded "Strongly disagree" to "Disagree" for the question, "The generated vocabulary included words which were redundant. All of the participants responded "Somewhat agree" to "Strongly agree" for the question "The vocabulary generated included words I would target during educational and/or speech therapy." Lastly, all of the participants responded "Somewhat agree" to "Strongly agree" for the question "Overall, the vocabulary generated was effective in helping me achieve targeted goals."}
    \caption{Participants reported that the automatically generated hotspots were relevant a majority of the time, and felt that they would be useful in achieving targeted goals in speech therapy.}
    \label{fig:likert-responses}
\end{figure*}

Through our post-questionnaire (see Appendix~\ref{sec:post-questionnaire}), we found that participants found that our prototype generated hotspots which they wanted to use, and did not frequently generate hotspots which they did not want to use. Additionally, our participants found that the generated hotspots were effective in supporting them in creating VSDs with hotspots they would use in educational or speech therapy and achieving the goals outlined in the case studies provided to them in the study (see Figure~\ref{fig:likert-responses}).

\subsubsection{Modifications Required for Prototype VSDs}

Our analysis of the generated hotspots aligned with our participants' responses. We found that, of the hotspots generated by our prototype (N = 178), 64.6\% of them were directly related to the scene, either about the people in the scene or the activity being depicted. However, we observed a high frequency of participants modifying generated hotspots to further personalize them to incorporate details provided to them in the case study (e.g., changing ``boy'' to the name provided in the case study).  The other trend we observed was participants modifying hotspots to make them more developmentally appropriate. Of the modifications made by participants (n = 8), 37.5\% made the hotspots more personalized and 37.5\% made them more developmentally appropriate. We also found that of the hotspots participants deleted (n = 57),  59.6\% were not relevant to the scene. Lastly, of the hotspots participants' added (N = 35), 79.41\% of them added personalized hotspots to the VSD.

\subsection{RQ3: Impact of Generated Hotspots}
VSDs configured with our prototype tended to be more homogeneous and participants tended to use more unique hotspots when using Snap Scene compared to our prototype. Additionally, participants seemed to rely heavily on the suggestions generated for them. We also observed some differences in the content across conditions. The parts-of-speech widely aligned across conditions, however, we saw a decrease in the use of hotspots for social interactions.

\subsubsection{Reliance on Generated Hotspots}
Of all of the hotspots generated by our prototype (N=178), 33.71\% were deleted, 4.49\% were edits, and 61.80\% were used unmodified. Generated hotspots composed 77.12\% of all hotspots (N = 153) used by participants when creating VSDs using our prototype.

\subsubsection{Homogeneity of VSDs Configured Across Conditions}

\begin{figure}
    \centering
    \includegraphics[width=0.6\linewidth]{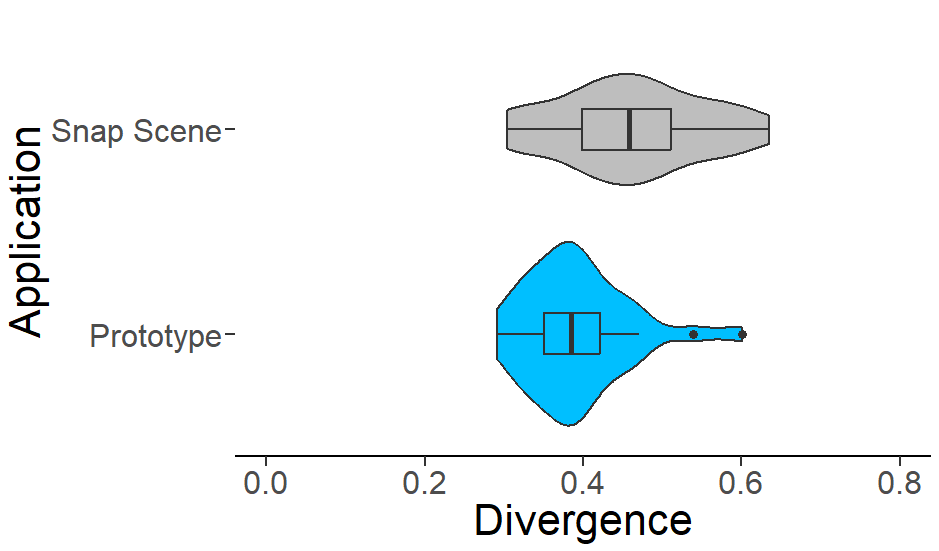}
    \Description[Stacked violin plot describing the homogeny of the VSDs configured with Snap Scene and our prototype.]{A stacked violin plot which describes the homogeneity of VSDs configured with Snap Scene (on the top, in gray) and our prototype (on the bottom, in blue). The Snap Scene VSDs were significantly less homogeneous than the VSDs made with our prototype. The median divergence (i.e., how different a VSD is from the "average" VSD) for Snap Scene and our prototype were 0.46 and 0.39, respectively. There were two outliers for the prototype divergence, approximately 0.55 and 0.6.}
    \caption{We found that VSDs that were configured with our prototype were more homogeneous than VSDs made with Snap Scene. The hotspots created with Snap Scene tended to diverge more from the average VSD based on semantic difference.}
    \label{fig:homogeneity}
\end{figure}

When participants configured VSDs with our prototype, the hotspots (i.e., communication options) were more similar to the average embedding of all hotspots created across both conditions using cosine similarity to the average embedding ($\bar{x} = 0.39 \pm 0.09$) compared to VSDs configured using Snap Scene ($\bar{x} = 0.46 \pm 0.07$). We found a statistically significant difference for the application used on the homogeneity of the hotspots created ($t(31) = 3.4037, p < 0.05$), with a medium effect size ($d = 0.60$), 95\% CI[0.26, 1.02], with higher homogeneity for VSDs configured using our prototype (see Figure~\ref{fig:homogeneity}).

\subsubsection{Unique Hotspots Used Across Conditions}

Across both conditions, 84 out of the total 247 hotspots were unique (34.01\%). In the Snap Scene condition 51 of 94 hotspots were unique (54.26\%). In the prototype condition 56 of 153 hotspots were unique (36.60\%).

\subsubsection{Hotspot Content Across Conditions}

Through our part-of-speech analysis we found that the content of the hotspots was largely similar across conditions. The hotspots were primarily composed of nouns and verbs in similar proportions in the Snap Scene and prototype conditions. The primary difference between the Snap Scene and prototype conditions was the inclusion of more advanced parts of speech (e.g., adjectives, pronouns, prepositions, adverbs, and articles) in the Snap Scene hotspots. See Table~\ref{tab:pos} for full description of the parts of speech used across conditions.


\begin{table}
\parbox{.45\linewidth}{
\centering
\begin{tabular}{@{}ccc@{}}
\toprule
\textbf{Part of Speech} & \textbf{Snap Scene} & \textbf{Prototype} \\ \midrule
Nouns & 64.0\% & 74.1\% \\
Verbs & 15.2\% & 15.7\% \\
Adjectives & 8.0\% & 5.4\% \\
Pronouns & 6.4\% & 4.2\% \\
Prepositions & 2.4\% & 0.0\% \\
Adverbs & 0.8\% & 0.0\% \\
Articles & 1.6\% & 0.0\% \\
Other & 1.6\% & 0.6\% \\ \bottomrule

\end{tabular}
\caption{Part of speech frequency for the Snap Scene and prototype conditions. The use of nouns, verbs, adjectives, and pronouns was similar across conditions. However, Snap Scene VSDs had more prepositions, adverbs, and articles.} 
\label{tab:pos}
}
\hfill
\parbox{.45\linewidth}{
\centering
\begin{tabular}{@{}ccc@{}}
\toprule
\textbf{Function} & \textbf{Snap Scene} & \textbf{Prototype} \\ \midrule
Information Transfer & 93.6\% & 98.7\% \\
Social Closeness & 5.3\% & 0.7\% \\
Expressing Wants \& Needs & 0.0\% & 0.0\% \\
Social Etiquette & 0.0\% & 0.0\% \\
Other & 1.1\% & 0.7\% \\ \bottomrule
\end{tabular}
\caption{The functions, as described by Light~\cite{light1988interaction}, of the hotspots used across conditions. We found that, across both conditions, participants' hotspots tended to focus on information transfer, however, there was substantially less focus on social closeness in the hotspots used when configuring VSDs with our prototype.}
\label{tab:functions}
}
\end{table}

Through our analysis of the function of the hotspots created across conditions, we found that, in both conditions, the hotspots participants used heavily focused on information transfer. In the Snap Scene condition, however, participants used more hotspots focused on social closeness than in the prototype condition. See Table~\ref{tab:functions} for a full description of the functions of the hotspots used across conditions.


\section{Discussion}

We conducted a within-subjects user study with pre-service SLPs to better understand how our genAI-enabled VSD application prototype, which automatically generates hotspots using a pre-trained multimodal LLM, affected users' efficiency and confidence in configuring VSDs. We also conducted analysis about how our prototype affected three quality metrics, relating to hotspot content, number of hotspots, and developmental appropriateness of hotspots. Our analysis also included a comparison of the content of hotspots across conditions and how much users relied on the automatically generated hotspots. In the following section, we discuss these results and the broader implications for our work.

\subsection{Support for Pre-Service SLPs Users in VSD Configuration}

SLPs possess a wealth of knowledge about linguistic and communication skill development, incorporating AAC interventions to aid in developing these skills, and personal information about their clients to use these interventions effectively. There are a lot of different kinds of interventions, and developing knowledge about their effective use and application may face barriers such as lack of access to experts~\cite{dejarnette2020preservice}, training opportunities~\cite{barman2023graduate, flores2025effect}, and educational opportunities~\cite{gohsman2023reported, thistle2023don}. We found that when participants, who were largely unfamiliar with VSDs prior to participating in our study, configured VSDs using our prototype they were faster and more confident in their VSDs. However, our findings regarding quality of the VSDs were mixed. The prototype performed better in one quality measure, specifically developmental appropriateness, but worse along others, specifically the number of hotspots used and the relevance of the hotspots to the scene. This indicates that while the use of a VSD creation application with automatically suggested hotspots is useful in increases users' confidence and efficiency, it is not a perfect solution. 

\subsubsection{Pre-service SLPs configure VSDs more quickly}
One of the main advantages of our prototype was its speed. Compared to Snap Scene, a current state-of-the-art tool, our prototype allowed participants to configure VSDs more quickly. This builds on prior work by Caron, Light, and Drager, who found that JIT programming of VSDs was more effective when there were fewer steps required~\cite{caron2016operational}. They demonstrated this with a prototype, EasyVSD, which is what Snap Scene was based on. Our prototype further reduces steps by automatically producing hotspot suggestions. With our prototype, SLPs can quickly evaluate these suggestions and add to them, modify, or delete them from the scene. By speeding up the process, SLPs and communication partners can quickly configure VSDs tailored to the immediate context which can improve rate of interaction, AAC user adoption, and increase the vocabulary with which AAC users interact~\cite{caron2016operational, light2019new}.  
Building upon work from Fontana de Vargas et al.~\cite{fontana2024co}, participants reported that the generated hotspots were largely relevant to the scene and useful in achieving the goals outlined in the case studies we provided them. Participants incorporated some or all of the automatically generated hotspots in most of the VSDs they configured ($31/32$ of VSDs configured with our prototype). 
Beyond our participants' perceptions; however, we observed a high incidence rate of the automatically generated hotspots being about objects in the background of the images or not relevant to the people or activities being depicted in the images.

\subsubsection{Pre-service SLPs are more confident in the VSDs they configure}
Participants also reported feeling significantly more confident about the VSDs they configured with our prototype compared to Snap Scene. This is important because when a user feels more confident in their ability to use an application, they are more likely to continue using that application. This phenomenon has been documented consistently across application domains, such as assistive technology~\cite{zolyomi2017technology, bandukda2021opportunities}, educational technology~\cite{staddon2023exploring, liu2025exploring}, and older adults' technology adoption~\cite{an2022understanding, moxley2022factors}. 
A lack of confidence has been linked to higher rates of abandonment of AAC devices with AAC users and their communication partners, often parents~\cite{mcnaughton2008child}. Our findings indicate that a system such as ours, may be beneficial in helping communication partners feel more confident using VSDs.
Providing initial suggestions of hotspots in our prototype to scaffold VSD configuration may have impacted participants' perceptions of the difficulty of the task and, in turn, made them feel more confident in configuring the VSDs. This has been observed in other contexts where JIT scaffolding has increased users' confidence~\cite{head2021augmenting, august2023paper, macneil2021framing}. 
Participants did not just feel more confident in their VSDs, they felt that they would be able to achieve the targeted goals of the case studies we presented to them. This feeling of self-efficacy in VSD use is promising, because, while the general feeling of self-efficacy and competence in AAC use of SLPs has increased in recent years~\cite{kovacs2021survey}, SLPs confidence in choosing the appropriate AAC devices for clients (i.e., feature matching) is not as high~\cite{beukelman2020augmentative}. Increasing SLPs confidence and general knowledge in VSDs may be a step towards increasing self-efficacy in feature matching.

\subsubsection{Pre-service SLPs used developmentally appropriate hotspots}

Participants created more single-word hotspots when using the prototype (91.5\%) than when using Snap Scene (78.7\%), which aligns with recommendations for VSD creation~\cite{holyfield2019programing, light2019new}. We also saw that when participants modified automatically generated hotspots (n = 8), they often changed hotspots to be more developmentally appropriate (37.5\% of the time). By using more single-word hotspots in VSDs, users are able to more flexibly combine a larger variety of words to communicate increasingly complex ideas, which is supportive of language development and acquisition~\cite{paul1997facilitating}.

\subsection{Negative Impacts of AI-Suggestions}

In the previous section, we outlined some of the positive effects of the automatic hotspot suggestions: faster configuration of VSDs, increased confidence, and using developmentally appropriate hotspots. However, these benefits come at a cost. 


%
\subsubsection{Configured VSDs were less aligned with best practices}

Despite these positive impacts, we also observed some more negative effects of genAI suggestions. In terms of quality impacts, we saw that participants tended to use significantly more hotspots when using our prototype and use hotspots which were not directly related to the scene. This conflicts with best practices from AAC research by potentially overwhelming the visual-cognitive processing capabilities of AAC users. 
Having too many hotspots can negatively impact linguistic development due to the likelihood for stimulus over-selectivity and overwhelming end users~\cite{dube2014potential, wilkinson2022judicious, hajjar2023visual}. By using hotspots which are not clearly referring to the people or activities within the scene, it circumvents one of the unique attributes of VSDs which makes them effective for language acquisition, incorporating a familiar context to anchor vocabulary~\cite{holyfield2019effect}. These results reflect the complexity of measuring quality in this context, but also raises concerns about just replacing VSD programming software with AI-enabled software, as it may lead to overwhelming visual cognitive processing. Lastly, we saw that when participants configured VSDs using our prototype, in almost every case (98.7\% of the time) the hotspots were meant to convey information and rarely develop social closeness (0.7\% of the time). When using Snap Scene, participants used proportionally more hotspots intended to develop social closeness (5.3\% of the time). VSDs have been shown to be effective in supporting social engagement between the VSD users and their communication partners, specifically their peers~\cite{laubscher2019effect, laubscher2022supporting}, so it is an important function to retain in VSDs. While this might be addressed by further refining the prompts for the genAI models, there currently exist misalignments between model behavior and best practices which need to be mitigated before focusing on aspects like prompt engineering. 
There may be interaction methods informed by prior HCI research which can be implemented to apply checks to ensure appropriateness and quality when configuring VSDs. One example of such an intervention could be to include prompting the users to reflect~\cite{haselager2024reflection} upon their created hotspots based on best practices and whether or not they have included personalized hotspots and hotspots for social closeness.

\subsubsection{Pre-service SLPs relied heavily on AI suggestions}

Through our analysis of how many of the generated hotspots were edited or deleted or when new hotspots were manually added, we saw that participants exhibited signs of over-reliance on the generated suggestions. Participants often accepted the hotspots suggested without any modifications (61.80\% of generated hotspots). Generated hotspots, included those which were edited made up 77.12\% of all of the hotspots created when using our prototype. Only 22.88\% of the total hotspots made using our prototype were manually added by our participants. 
This could be related to participants having a false sense of confidence in the genAI to generate high-quality suggestions~\cite{kelly2023capturing}. Although our intention was for participants' to use the generated hotspots as a starting point for their VSD creation, it is not unexpected that they over-relied on the automated hotspots. This commonly occurs in genAI-powered systems used by non-experts for complex tasks~\cite{bucina2021to, prather2024widening}. Over-reliance on AI is a well-studied phenomenon, and there are multiple approaches which have been validated to combat over-reliance. A potential solution could be to implement a validation step which checks a user's configured VSD against best practices and explains~\cite{vasconcelos2023explanations} what changes should be made.

\subsubsection{Communication options were more homogenous}

The last and potentially most impactful negative impact we saw observed was the increased homogeneity and reduction of unique hotspots of VSDs configured with our prototype. Increased homogeneity when using genAI-powered systems has been observed in the context of using genAI as a creativity support tool~\cite{anderson2024homogenization}. This homogenizing effect has been observed in other contexts as well, including writing~\cite{bauer2025does, agarwal2025ai}, crowdworker output~\cite{veselovsky2025prevalence}, and people's opinions and beliefs~\cite{durmus2023towards, jakesch2023cowriting, weidinger2021ethical}. As previously mentioned, the reason why VSDs are engaging and effective for users is that they are personally relevant, and just-in-time programming has enabled communication partners to specifically tailor the VSDs for users~\cite{drager2019aac, holyfield2019effect, chapin2022effects}, sometimes with input from the VSD users themselves~\cite{holyfield2017typical}. Evidence of homogenization of VSDs is potentially harmful, as homogenized VSDs represent a narrower range of options that can be communicated. It is important to offer a diverse range of options to support broader expression via VSDs. The goal  is to expand communication, and it is problematic if integrating AI into VSDs inadvertently limits it. The inadvertent homogenization effect of genAI has lead to many researchers to call for solutions to mitigate this homogenization~\cite{agarwal2025ai, anderson2024homogenization}, and more research needs to be done to understand how to apply these potential solutions to AI-enabled AAC devices.

\subsection{Addressing the Risks of AI in AAC Devices}

In the previous sections we discussed some of the positive and negative findings of using genAI in our VSD configuration prototype. While there were some positive impacts of using our prototype - increased efficiency, confidence, and the use of developmentally appropriate hotspots - 
there are important negative impacts, such as over-reliance and homogenization which are known in design and education contexts, but have been overlooked in previous studies of AI-enabled AAC devices~\cite{valencia2024compa, fontana2024co}. 

Because of these newly discovered issues, this prototype could be potentially useful in an educational context for scaffolding VSD creation rather than requiring pre-service SLPs to start from scratch. Training is a really important part of self-efficacy and the effective use of AAC devices~\cite{flores2025effect,conlon2024confidence}. Introducing this as a training tool could make it easier for pre-service SLPs to develop confidence and familiarity in using VSDs while helping professionals and experts to train and correct pre-service SLPs with some of the best practices we have discussed. But the same problems remain for in many AI systems: over-reliance~\cite{zastudil2023generative, hou2024effects, fui2023generative}, homogenization~\cite{bauer2025does, agarwal2025ai, durmus2023towards, jakesch2023cowriting, weidinger2021ethical}, and a lack of critical engagement~\cite{nakic2024chatgpt, van2023chatgpt, prather2024widening}.


Researchers have previously advocated for the integration of AI into AAC by presenting positive empirical outcomes~\cite{valencia2024compa, fontana2024co}. However, our findings offer a more nuanced perspective that reiterates some of the positive aspects while also identifying potential pitfalls.
We identified a potential risk of over-reliance on AI suggestions where participants passively accepted suggestions without modifying them. We also identified tradeoffs between efficiency and quality. Participants were more confident when using AI suggestions in our study despite mixed findings regarding the quality of VSDs generated with AI support.  Trusting that the suggestions are good when they are not, in the case of AAC, can significantly harm users' linguistic and social development. By providing communication options that are not related to scene or by overwhelming users with too many communication options, it reduces the usefulness of AAC devices~\cite{baxter2012barriers}, which can also risk the abandonment of AAC devices entirely~\cite{baxter2012barriers}. 

Finally, we observed that the hotspots created with AI assistance tended to be more homogeneous. Given that AAC devices are most effective when carefully personalized to the user, this homogenization is problematic because it essentially represents sensible defaults. The homogenizing effect of genAI that we observed presents a disturbing trend - the creation of VSDs and potentially other AAC displays which are becoming more and more similar. The benefit of JIT programming of AAC displays is that they can be incredibly personalized to each user~\cite{holyfield2019programing}. Using JIT programming to configure less unique AAC displays results in a display that is more similar to a default display, which are not as useful to anyone.

\section{Limitations and Future Work}

While the core functionality (i.e., taking or uploading an image, manual creation and editing of hotspots) of the two application used in our study was the same, there are minor differences (e.g., icons, fonts) which may impact the usability of our prototype as compared to Snap Scene. Additionally, our user study only included one group of communication partners, pre-service SLPs. Future work could also include other communication partners such as parents, caregivers, and friends. We hypothesize that the trends of over-reliance and homogenization would be more apparent with these communication partners, as pre-service SLPs contain specific knowledge about how language and communication skills are developed. 

Our prototype represents a lower bound for how AI might be integrated into AAC, specifically VSDs. With more carefully crafted guardrails, some of the issues we identified in this study (e.g., over-reliance, homogeniety) might be mitigated. For example, through more advanced prompting techniques (e.g., self-critique~\cite{madaan2023self}), homogeneity could be reduced; however, given the propensity for over-reliance, this approach gives more agency to the AI. Future work is necessary to refine our prototype and add in interactive features that would better support the configuration of high-quality VSDs.



\section{Conclusion}
VSDs have been shown to be effective for many communicators who use AAC devices; however, there are many barriers which prevent their widespread adoption outside of clinical contexts. The primary barrier is that they are difficult to configure for communication partners who are not experts in VSDs and JIT configuration. We developed a prototype VSD which uses genAI to scaffold the configuration process. We evaluated this prototype with 16 pre-service SLPs with mixed results. We found that while our prototype increases users' confidence and self-efficacy, we see substantial negative impacts related to over-reliance on AI suggestions, divergence from best practices in VSD creation, and homogenization of VSDs configured with the support of AI.

\begin{acks}
This work was partially funded by the Temple University Pervasive Computing REU Site and Convergence Accelerator Grant (National Science Foundation grant numbers CNS-2150152 and ITE-2236352).
\end{acks}

\bibliographystyle{ACM-Reference-Format}
\bibliography{sample-base}

\appendix
\section{Images and Case Studies Used for VSD Creation}
\label{sec:all-prompts}

\begin{figure*}
 \begin{subfigure}{0.49\textwidth}
     \includegraphics[width=\textwidth]{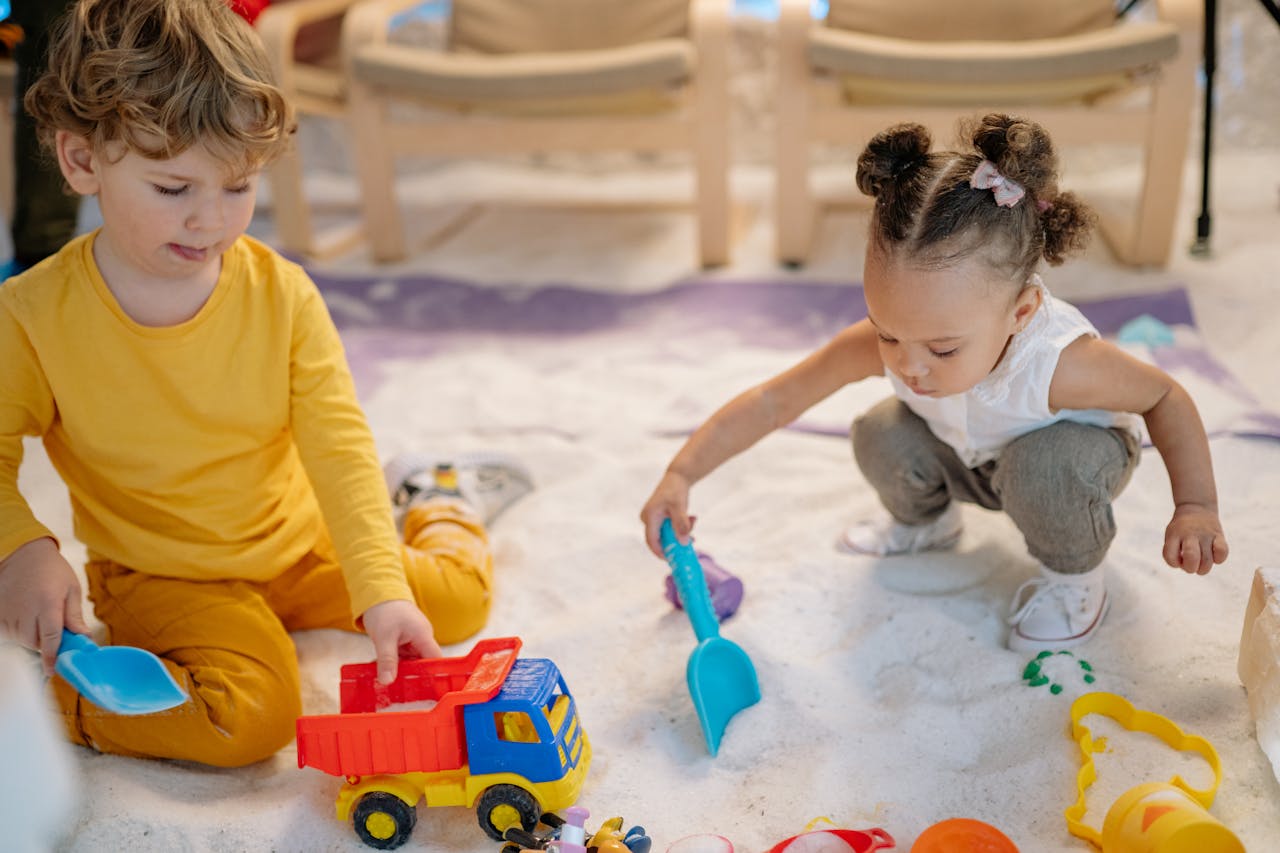}
     \caption{VSD 1. Children playing in a sandbox.}
     \label{fig:sandbox-vsd}
     \Description{Two children playing in a sandbox. they are playing with a dump truck, sand shovels, and various other sand toys.}
 \end{subfigure}
 \hfill
 \begin{subfigure}{0.49\textwidth}
     \includegraphics[width=\textwidth]{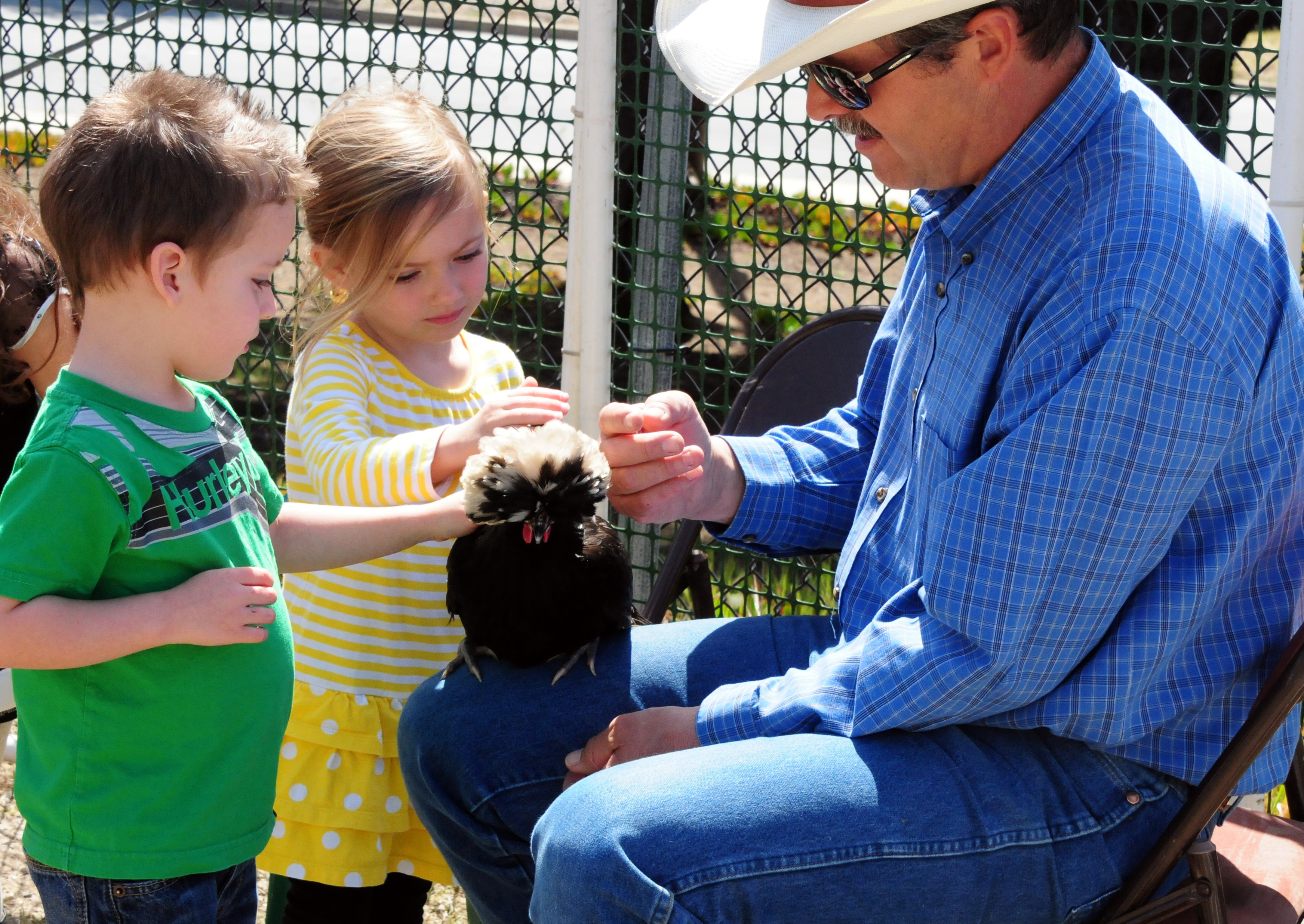}
     \caption{VSD 2. Children petting a chicken.}
     \label{fig:chicken-vsd}
     \Description{Two children petting a chicken which is sitting on a man's lap. The man is wearing a cowboy hat, and the children look happy.}
 \end{subfigure}
 
 \medskip
 \begin{subfigure}{0.49\textwidth}
     \includegraphics[width=\textwidth]{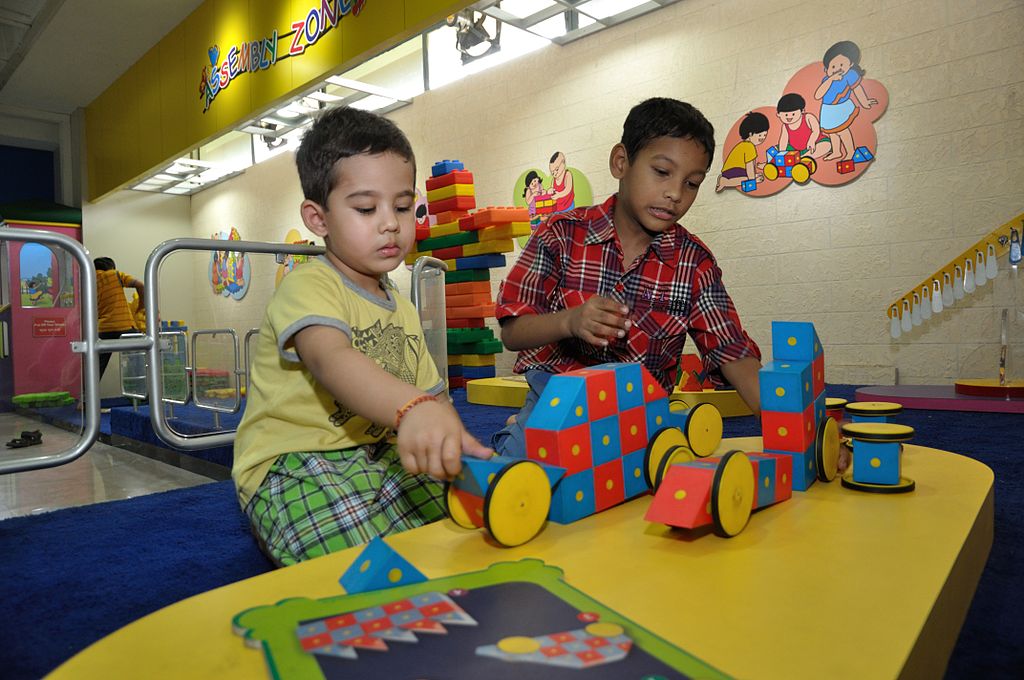}
     \caption{VSD 3. Children playing with blocks at a table.}
     \label{fig:blocks-vsd}
     \Description{Two children playing with blocks at a table at a play center. They look to be building a vehicle together using the blocks.}
 \end{subfigure}
 \hfill
 \begin{subfigure}{0.49\textwidth}
     \includegraphics[width=\textwidth]{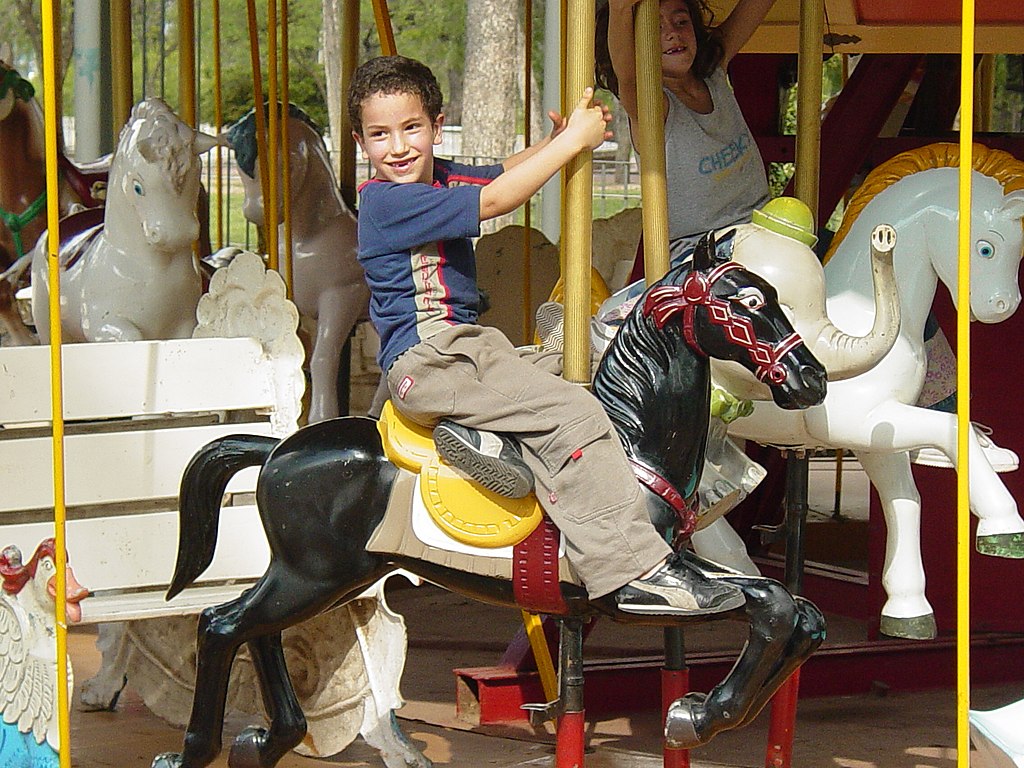}
     \caption{VSD 4. Children riding a carousel.}
     \label{fig:carousel-vsd}
     \Description{Two children riding a carousel on black and white horses. They look very excited to be on the ride.}
 \end{subfigure}

 \caption{All four VSD images we used in our user study.}
 \label{Label}

\end{figure*}

\textbf{VSD 1. Children Playing in a Sandbox.} (see Figure~\ref{fig:sandbox-vsd}) ``Imagine you took this photo to create a VSD for the child in the yellow shirt. He is a 7 year old on the autism spectrum. His goals in therapy are focused on building an expressive and receptive single word vocabulary, as his parents and professionals estimate that he understands about 25 words and spontaneously communicates about 10 words. He loves all forms of vehicles and playing with his sister Cara, also pictured.''

\textbf{VSD 2. Children Petting a Chicken.} (see Figure~\ref{fig:chicken-vsd}) ``Imagine the parent of the girl in the yellow shirt in this photo sent the photo to you to create a VSD of a fun moment from the weekend. She is a 4 year old on the autism spectrum. Her goals in therapy are focused on expanding her receptive and expressive single word vocabulary (estimated around 150 words) and the emergence of two-word combinations. She loves all animals and spending time with her friend Dean, also pictured.''

\textbf{VSD 3. Children Playing with Blocks at a Table.} (see Figure~\ref{fig:blocks-vsd}) ``Imagine you took this photo to create a VSD for the child in the red plaid shirt. He is a 9 year old on the autism spectrum. His goals in therapy are focused on building a single word vocabulary because his estimated receptive word knowledge to be fewer than 50 words and his spontaneous use of words expressively is limited (about five words). He loves colorful toys and all building-based toys. He also enjoys playing with his little cousin Jonas, also pictured.''

\textbf{VSD 4. Children Riding a Carousel.} (see Figure~\ref{fig:carousel-vsd}) ``Imagine the parent of the girl in the white shirt in this photo sent the photo to you to create a VSD of a fun moment from the weekend. She is a 6 year old on the autism spectrum. Her goals in therapy are focused on expanding her single word vocabulary (expressive and receptive) as she is estimated to understand about 250 words and use about 50 spontaneously. Her therapy is also focused on the emergence of two-word combinations. She loves animals and spending time with her big brother, Sam, also pictured.''

\section{Post-Questionnaire}
\label{sec:post-questionnaire}
Participants were asked to rate their agreement to the following statements at the conclusion of our user study. Each statement was accompanied by a 7-point Likert scale for measuring agreement (Strongly agree, Agree, Somewhat agree, Neither agree or disagree, Somewhat disagree, Disagree, Strongly disagree).
\begin{enumerate}
    \item The generated vocabulary included words I wanted to use.
    \item The generated vocabulary included words I did not want to use.
    \item The generated vocabulary included words I would not have thought of that are relevant.
    \item The generated vocabulary included words which were redundant.
    \item The vocabulary generated included words I would target during educational and/or speech therapy.
    \item Overall, the vocabulary generated was effective in helping me achieve targeted goals.
\end{enumerate}

\end{document}